\documentclass[conference]{IEEEtran}
\usepackage{lmodern}
\usepackage{graphicx}
\usepackage{mathrsfs}
\usepackage{tikz}
\usepackage{booktabs}
\usepackage{epstopdf}
\usepackage{enumerate}
\usepackage{psfrag}	
\usepackage{array}
\usepackage{multirow,hhline}
\usepackage{exscale}
\usepackage{color}
\usepackage{epsfig,subfigure}
\usepackage{cite}
\usepackage{amsthm}
\usepackage{nicefrac}
\usepackage{standard}

\usepackage[margin=54pt]{geometry}

\theoremstyle{definition}
\newtheorem{theorem}{Theorem}
\newtheorem{remark}{Remark}

\definecolor{c1}{rgb}{0,0,0} 
\definecolor{c2}{rgb}{0,0,0}
\definecolor{c3}{rgb}{0,0,0} 
\definecolor{c4}{rgb}{0,0,0}
\definecolor{c5}{rgb}{0,0,0}
\definecolor{c6}{rgb}{0,0,0}
\definecolor{c7}{rgb}{0,0,0}
\definecolor{c8}{rgb}{0,0,0}

\begin{document}
\IEEEoverridecommandlockouts
\title{A Systematic Approach for Interference Alignment in CSIT-less Relay-Aided X-Networks\thanks{This work is supported in part by the German Research Foundation, Deutsche
Forschungsgemeinschaft (DFG), Germany, under grant Li 659/13.}} 
\author{
\IEEEauthorblockN{Daniel Frank, Karlheinz Ochs, and Aydin Sezgin}
\IEEEauthorblockA{Chair of Communication Systems\\
RUB, Germany\\
Email: { \{daniel.frank, karlheinz.ochs, aydin.sezgin\}@rub.de}}
}
\maketitle

\begin{abstract}
The degrees of freedom (DoF) of an X-network with M transmit and N receive nodes utilizing interference alignment with the support of $J$ relays each equipped with $L_j$ antennas operating in a half-duplex non-regenerative mode is investigated. Conditions on the feasibility of interference alignment are derived using a proper transmit strategy and a structured approach based on a Kronecker-product representation. The advantages of this approach are twofold: First, it extends existing results on the achievable DoF to generalized antenna configurations. Second, it unifies the analysis for time-varying and constant channels and provides valuable insights and interconnections between the two channel models. It turns out that a DoF of $\nicefrac{NM}{M+N-1}$ is feasible whenever the sum of the $L_j^2 \geq [N-1][M-1]$.
\end{abstract}

\section{Introduction}\label{sec1}
Interference alignment is a powerful tool to boost the rates of users in an interference limited communication environment \cite{maddah,2,3,1}.
Initially it was utilized for the multiple antenna X-channel and the $K$-user interference channel, showing that the sum capacity of those networks can be approximately characterized as a function of the signal-to-noise-power ratio $\mathsf{SNR}$ by
\begin{align*}
C_\Sigma= \mathrm{DoF}\log\left(\mathsf{SNR}\right)+o\left(\log(\mathsf{SNR})\right).
\end{align*}
For asymptotically high $\mathsf{SNR}$, the second term vanishes while the first term dominates the behavior in that regime. Here, the pre-log term is referred to as degrees of freedom (DoF) and is interpreted as the number of parallel interference-free point-to-point communication links inherent in the network under investigation.

It is important to note that most of the work discussed so far assumed instant and global channel state information (CSI) at all nodes to effectively implement interference alignment, with a channel that is fast fading. Missing global CSI at the transmitters (CSIT) leads to loss of DoF for many networks \cite{7,8}. Furthermore, the DoF is obtained by letting the number of channel uses be arbitrarily high.

Introducing relays to a wireless network is helpful to improve the achievable rates \cite{22,25}. In this context, highly relevant is the work in \cite{29}, where it was shown that relaying, feedback, and cooperation do not increase the DoF for fully connected networks. However, relays can be used to transform a static channel into an equivalent time varying channel in order to achieve optimal DoF using interference alignment in static environments \cite{30,31,32}. Besides, \cite{33} presents a relay-aided transmission scheme which results in an optimal DoF for the $K$-user interference channel with finite channel uses.

More recently it was shown in \cite{tian3} that adding half-duplex relays with global CSI to a multi-user X-network is sufficient to achieve the optimal DoF with finite channel uses when the transmitters have no CSI. Unfortunately, in \cite{tian3} a symmetric antenna configuration at the relays is needed and it is crucial that the channel is time varying. From a practical point of view, it would be more useful to support asymmetric relay antenna configurations. Furthermore, global CSI at the relays is a hard to guarantee condition for fast fading channels and is more agreeable to ensure for constant channels.

In this paper, we utilize the idea of \cite{tian3}. However, we use a structured approach based on the Kronecker-product representation. This extends the results on the achievable DoF of  \cite{tian3} to generalized antenna configurations at the relays. In addition, we use a transmit strategy which assures that our approach is applicable in static environments as well.

The rest of the paper is organized as follows. In Section \ref{sec2}, we introduce the relay-aided $[M \times N]$-user X-network and describe the system model. The main result of the paper is given in Section \ref{sec3}, where we present a relay-aided alignment scheme which achieves the optimal DoF of both the time varying and static multi-user X-network without CSIT. Finally, we conclude our results in Section \ref{sec4}.

\section{System model}\label{sec2}
\begin{figure}[ht]
 \centering
     \includegraphics[width=.48\textwidth]{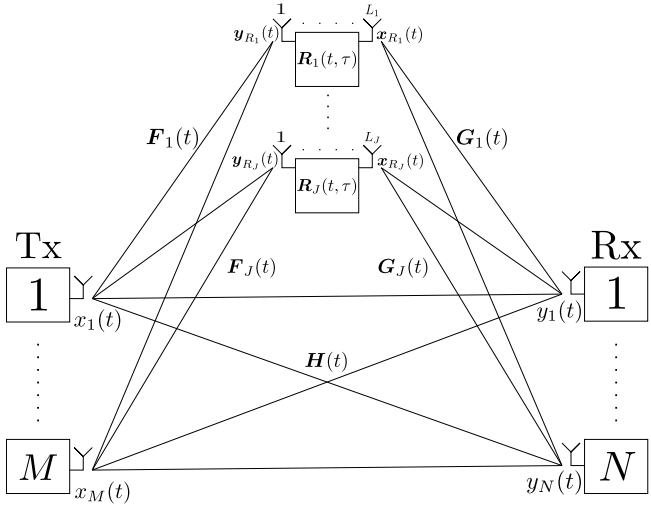}
    \caption{Relay-aided communication in an $[M\times N]$-user X-network}
    \label{fig:X-channel}
\end{figure}

We consider a relay-aided communication in an $[M\times N]$-user X-network as depicted in Figure \ref{fig:X-channel}. Here, we have $M$ transmitters and $N$ receivers each equipped with one antenna. It is assumed that the transmitters have no CSI whereas the receivers have global CSI.
In this network, transmitter (Tx) $m$, $1 \leq m \leq M$, has a message $w_{nm}$ for receiver (Rx) $n$, $1 \leq n \leq N$, and therefore $NM$ messages are communicated in total. The message $w_{nm}$ is encoded into a codeword. We denote one symbol of this codeword as $d_{nm}$ which is used to construct the complex transmit signal $x_m(t) \in \setC$ at Tx $m$, where the index $t$ indicates the time slot of the transmission. The average transmit power for each $x_m(t)$ is given by $P$. The received signal at Rx $n$ is denoted as $y_n(t) \in \setC$ and is disturbed by the noise $z_n(t)$. That is, we assume an additive white Gaussian noise (AWGN) with zero mean and unit variance.

The communication is supported by $J$ half-duplex relays, where the $j$-th relay is equipped with $L_j$ antennas and has global CSI. We combine the $L_j$ transmit signals and the $L_j$ received signals at the $j$-th relay to the complex signal vectors $\vect x_{R_j}(t) \in \setC^{L_j}$ and $\vect y_{R_j}(t) \in \setC^{L_j}$, respectively. Moreover, the average transmit power is given by $P$ and the received signal vector is corrupted by the noise vector $\vect z_j(t)$, where we assume an AWGN with zero mean and unit covariance matrix.

To have a more condensed formulation of the system model, we stack all transmit signals $x_m(t)$ and all received signals $y_n(t)$ into the vectors $\vect x(t) \in \setC^{M}$ and $\vect y(t) \in \setC^{N}$, respectively. When we do the same for all $z_n(t)$, the input-output relation reads:
\begin{subequations}
    \begin{align}
      \vect y(t) &= \mat H(t) \vect x(t) + \dsum_{j=1}^{J}{\mat G_j(t) \vect x_{R_j}(t)} + \vect z(t)~, \label{eqn:recv_signals_a}  \\
       \vect y_{R_j}(t) &= \mat F_j(t)\vect  x(t) + \vect z_j(t)~. \label{eqn:recv_signals_b}
    \end{align}
\end{subequations}
The matrix $\mat H(t) \in \setC^{N \times M}$ is the channel matrix between the transmitters and the receivers and the complex entry $h_{nm}(t) \in \setC$ of $\mat H(t)$ is the channel coefficient between Tx $m$ and Rx $n$. Moreover, the channel between the transmitters and the $j$-th relay is represented by $\mat F_j(t) \in \setC^{L_j \times M}$. We denote the channel vector between Tx $m$ and relay $j$ as $\vect f_{jm}(t) \in \setC^{L_j}$ which is the $m$-th column of the matrix $\mat F_j(t)$. The channel between relay $j$ and the receivers is indicated by $\mat G_j(t) \in \setC^{N \times L_j}$. Furthermore, the channel vector between relay $j$ and Rx $n$ is represented by $\mt{\vect g}_{nj}(t) \in \setC^{1 \times L_j}$ which is the $n$-th row of the matrix $\mat G_j(t)$. Here, $\mt{(\cdot)}$ denotes the transpose operator. We assume that all channel coefficients are independently drawn from continuous distributions. For this reason, the channel matrices $\mat H(t) \in \setC^{N \times M}$, $\mat F_j(t) \in \setC^{L_j \times M}$ and $\mat G_j(t) \in \setC^{N \times L_j}$ are full rank almost surely.

The focus of this work is the high-$\mathsf{SNR}$ behavior of this system utilizing the DoF metric \cite{jafar}. To this end, let $R_{nm}(P)$ be the rate with which the message $w_{nm}$ is communicated. We define $\mathcal{C}$ as the set of all achievable rate tuples $[R_{11},\hdots, R_{NM}]$. Then, the DoF metric is defined as
\begin{align}
 \text{DoF} = \lim\limits_{\mathsf{SNR} \rightarrow \infty} C_\Sigma(\mathsf{SNR})/\text{log}(\mathsf{SNR})~.
\end{align}
Here, $C_\Sigma(\mathsf{SNR}) = \underset{\mathcal{C}}{\text{max}}~R_\Sigma $ is the sum capacity of the network which is defined as the maximum achievable sum rate $R_\Sigma = \tsum_{n,m}{R_{nm}}$. Since we are considering the DoF of the channel, we neglect the noise terms in equations (\ref{eqn:recv_signals_a})-(\ref{eqn:recv_signals_b}) in the sequel.

\section{$[M \times N]$-User X Channel with multiple antenna relays}\label{sec3}
In this section, we present a relay-aided transmission scheme for the $[M \times N]$-user X-network which achieves the maximum DoF $\nicefrac{NM}{M+N-1}$ \cite{1}. The following theorem summarizes the main result.
\begin{theorem}
Consider an $[M\times N]$-user X-network with $J$ relays, where the $j$-th relay is equipped with $L_j$ antennas. Suppose that the transmitters have no CSI, while the relays and receivers have global CSI. Then, the DoF $\nicefrac{NM}{M+N-1}$ is feasible whenever the sufficient condition 
\begin{align}
L = \dsum_{j=1}^{J}{L_{j}^{2}} \geq [N-1][M-1] 
\label{alignment_condition}
\end{align}
is fulfilled.
\label{thm:theorem1}
\end{theorem}
The general idea to prove the DoF $\nicefrac{NM}{M+N-1}$ is that we need to communicate $NM$ symbols reliably in $T = [M+N-1]$ time slots, where each receiver is able to decode its $M$ desired symbols. Therefore, we use a $T$-symbol extended channel where the transmit and received signals become vectors of length $T$. The proof including the communication strategy which achieves the DoF of the network is described in the following subsections in more detail.
\begin{figure*}
  \begin{align}
         \tilde{\mat H} &=
         \mleft{cccccccc}
         \mat H(1) & \zero & \cdots & \zero &\zero & \zero & \cdots & \zero \\
         \zero & \mat H(2) & \cdots & \zero &\zero & \zero & \cdots & \zero \\
         \vdots & \vdots & \vdots  & \vdots &\vdots & \vdots & \vdots & \vdots \\
         \zero & \cdots & \cdots &   \mat H(N) &\zero & \zero  & \cdots & \zero \\
         \mat S(N+1,1) &  \mat S(N+1,2) &  \cdots  & \mat S(N+1,N) &  \mat H(N+1) & \zero & \cdots &\zero \\
         \mat S(N+2,1) &  \mat S(N+2,2) &  \cdots  & \mat S(N+2,N) &  \zero & \mat H(N+2) & \cdots &\zero \\
         \vdots & \vdots  & \vdots & \vdots & \vdots & \vdots & \vdots  & \vdots\\
         \mat S(T,1) &  \mat S(T,2) &\cdots  & \mat S(T,N)  & \zero  &  \zero & \cdots &\mat H(T) \\
         \mright
  \label{eqn:X_matrix}
  \end{align}
 \hrule
\end{figure*}
\subsection{Transmitters and Relays}
In the first $N$ time slots, all transmitters are active and send their symbols to the receivers, while the relays just listen and store the received signals. In each of the following $[M-1]$ time slots, all relays and only one specific transmitter (Tx $[t-N]$, $t \in \{[N+1],\ldots, T\}$) are active to support the communication. In more detail, we have for the transmit signal vector at the transmitters
\begin{align}
 \vect x(t) &=
 	\begin{cases}
 	\vect d_t
 	&,1\leq t\leq N\\
    \vect e_{m'}^{M} \mt{(\vect e_{m'}^{M})} \tsum_{\tau =
 1}^{N}{\vect d_{\tau}}&,N<t\leq T,
     \end{cases}
 \label{eqn:X_transmitters}
 \end{align}
where
\begin{align}\label{eqn:X_transmitters2}
m'=[t-N]
\end{align} 
 and $\vect e_{k}^{K}$ is the $k$-th column of the $K \times K$ identity matrix $\unit_K$. Here, the vector $\vect d_t = \mt{\left[d_{t1},\hdots, d_{tM} \right]}$ contains all the symbols which are desired at Rx $t$. For the transmit signal vector at the $j$-th relay, we have
\begin{equation}
   \begin{aligned}
    \vect x_{R_j}(t) =
     \begin{cases}
      \zero_{L_j}&,~1\leq t\leq N\\
      \tsum_{\tau = 1}^{N}{\mat R_{j}(t,\tau) \vect y_{R_j}(\tau)}&,~N<t\leq T,
     \end{cases}
  \end{aligned}
\label{eqn:X_relays}
\end{equation}
where $\zero_K$ is a zero column vector with dimension $K$ and $\mat R_j(t,\tau) \in \setC^{L_j \times L_j}$ is a precoding matrix.
From equations (\ref{eqn:X_transmitters}) and (\ref{eqn:X_relays}), we can see that in time slot $1\leq t\leq N$, all transmitters send their symbol desired at Rx $n = t$ and the relays remain silent. 
In the time slots $[N+1]\leq t\leq T$, the relays and only the $m'$-th transmitter are active to support the communication. The relays apply a precoding matrix $\mat R_j(t,\tau)$ to the signals which they have received in the time slots $\tau = 1,\hdots,N$ and send the sum of the precoded signals to the receivers. The $m'$-th transmitter performs joint-beamforming with the relays by sending a linear combination of its symbols. As we will see, this transmit strategy at the transmitters is crucial to make the scheme work for time constant channels.
\subsection{Received Signals} 
From the first $N$ time slots, Rx $n$ gets one linear combination of its $M$ desired symbols and $[N-1]$ linear combinations of undesired symbols. Therefore, Rx $n$ is not able to decode its desired symbols. The relay-aided transmission in the following $[M-1]$ time slots provides each receiver with $T$ signal dimensions. Since we need $M$ dimensions for interference free data transmission, the relays have to choose their precoding matrices $\mat R_j(t,\tau)$ in such a way that all interference signals are aligned into the remaining $[N-1]$ dimensional space. To this end, initially we state the time extended input-output relation by considering the transmit strategy of the relays and the equations (\ref{eqn:recv_signals_a})-(\ref{eqn:recv_signals_b}) as
\begin{align}
\vect y = \tilde{\mat H}\vect x
\label{eqn:received signal_vector}
\end{align}
with $\vect x=\text{vec}([\vect x(1), \hdots, \vect x(T)]) \in \setC^{TM}$ and $\vect y=\text{vec}([\vect y(1), \hdots, \vect y(T)]) \in \setC^{TN}$. Here, $\text{vec}(\mat A)$ denotes the vec-operator which creates a column vector from a matrix $\mat A$ by stacking the columns of $\mat A$ one below the other.
The matrix $\tilde{\mat H} \in \setC^{TN \times TM}$ which is a lower triangular channel matrix, is given in (\ref{eqn:X_matrix}) on the top of the page, where 
\begin{align}\label{matrix_s}
\mat S(t,\tau) &= \dsum_{j=1}^{J}{\mat G_j(t) \mat R_j(t,\tau) \mat F_j(\tau)}~.
\end{align}
Multiplying (\ref{eqn:received signal_vector}) from the left by $[\unit_T \otimes \mt{(\vect e_n^N)}]$, gives us the received signal vector $\vect y_n \in \setC^{T}$ at Rx $n$, where $\otimes$ is the Kronecker operator, i.e.,
 \begin{align}
 \vect y_n = [\unit_T \otimes\mt{(\vect e_n^N)}] \tilde{\mat H}\vect x~.
 \label{eqn:received signal_vector_usern1}
 \end{align}
 Using the relation $\vect x(t) = \tsum_{m=1}^{M}{\vect e_m^M \mt{(\vect e_m^M)} \vect x(t)}$, equation (\ref{eqn:X_transmitters}) can be expressed as 
\begin{align}
\vect x(t) &=
\begin{cases}
\tsum_{m=1}^{M}{d_{tm} \vect e_m^{M}}&,1\leq t\leq N\\
\vect e_{m'}^{M}\tsum_{\tau=1}^{N}{\tsum_{m=1}^{M}{d_{\tau m}
\delta_{m-m'}}}&,N<t\leq T,
\end{cases}
\label{eqn:X_transmitters3}
\end{align}
where 
\begin{align*}
\delta_{m-m'} = 
\begin{cases}
1,~&m=m'\\
0,~&m \neq m'
\end{cases}
\end{align*}
is the Kronecker delta and $m'$ is defined as in (\ref{eqn:X_transmitters2}). 
Substituting (\ref{eqn:X_transmitters3}) into (\ref{eqn:received signal_vector_usern1}), we are able to formulate $\vect y_n$ as
\begin{subequations}
\begin{align}
      \vect y_n =& \dsum_{\tau=1}^{N}{\dsum_{m=1}^{M}{d_{nm}h_{nm}(\tau) \hat{\vect h}_{nm}(\tau)}}
\label{eqn:received signal_vector_usern2}
\end{align}
with
\begin{align}
\hat{\vect h}_{nm}(\tau) =   
           \mleft{c}
           \vect e_\tau^N\\
           \frac{s_{nm}(N+1,\tau)+h_{n1}(N+1)\delta_{m-1}}{h_{nm}(\tau)}\\
           \vdots\\
           \frac{s_{nm}(T,\tau)+h_{n[M-1]}(T)\delta_{m-[M-1]}}{h_{nm}(\tau)}\\
           \mright
\label{eqn:received signal_vector_usern3}
\end{align}
and
\begin{align}
s_{nm}(t,\tau) = \dsum_{j=1}^{J}{\mt{\vect g}_{nj}(t) \mat R_j(t,\tau) \vect f_{jm}(\tau)}~.
\label{eqn:entry_s}
\end{align}
\end{subequations}
\begin{figure*}[ht]   
 \centering
    \includegraphics[width=1\textwidth]{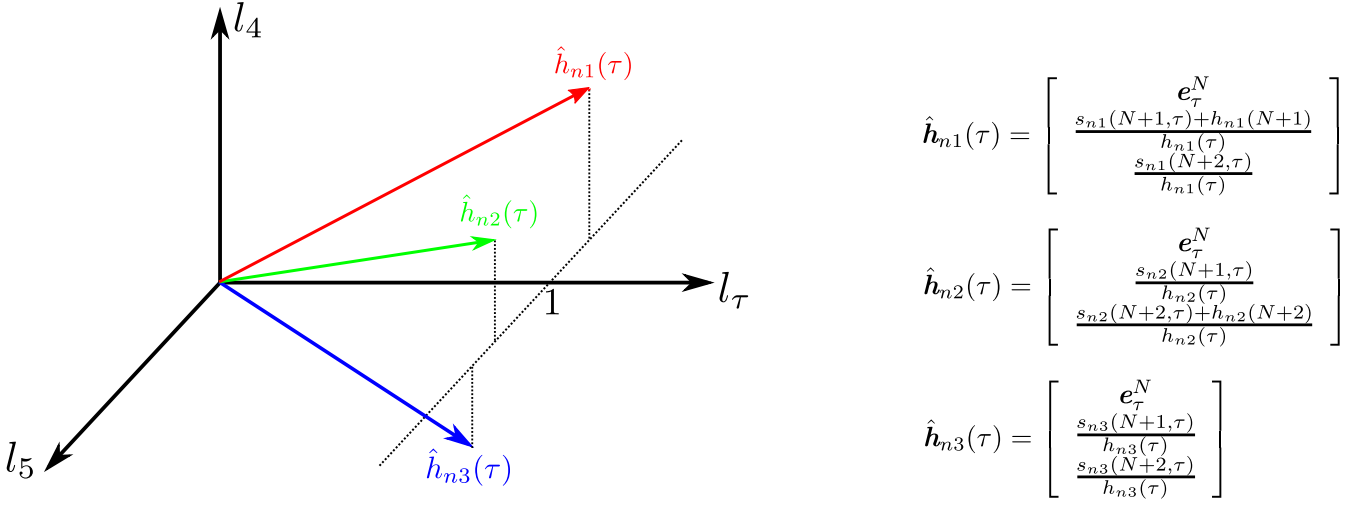}
    \caption{Structure of the vectors $\hat{\vect h}_{nm}(\tau)$ for $M=3$ transmitters. Since all vectors have $l_\tau = 1$, we are only able to force the vectors into a one dimensional space by setting each of $l_4$ and $l_5$ to be equal for all three vectors.}
    \label{fig:alignment}
\end{figure*}

\subsection{Interference Alignment}
From equation (\ref{eqn:received signal_vector_usern2}), we get $NM$ vectors at Rx $n$. Here, the $M$ vectors for $n=\tau$ are desired at Rx $n$ since all intended symbols for Rx $n$ are transmitted in time slot $\tau = n$ (see (\ref{eqn:X_transmitters})). Hence, we get $M$ desired vectors and $[M(N-1)]$ interference vectors at each receiver:
\begin{align}
\begin{split}
      \vect y_n =& \underbrace{\dsum_{m=1}^{M}{d_{nm}h_{nm}(n) \hat{\vect h}_{nm}(n)}}_{\text{desired}} \\
      +& \underbrace{\dsum_{\substack{\tau=1 \\ \tau \neq n}}^{N}{\dsum_{m=1}^{M}{d_{\tau m}h_{nm}(\tau) \hat{\vect h}_{nm}(\tau)}}}_{\text{undesired}}~.
 \end{split}
\label{eqn:received signal_vector_usern4}
\end{align}
We aim to align the interference vectors into an $[N-1]$ dimensional space.
Notice that the structure of the first part ($\vect e_\tau^N$) of the vectors $\hat{\vect h}_{nm}(\tau)$ is the same for all $m \in \{1,\ldots, M\}$ for a given $\tau \in \{1,\ldots,N  \}$ and varies for different $\tau$. In order to achieve our alignment goal, we have to align the $M$ interference vectors $\hat{\vect h}_{nm}(\tau)$ for a given $\tau \neq n$ into a one dimensional space. This forces the $[M(N-1)]$ interference vectors into an $[N-1]$ dimensional space and leaves $M$ dimensions for the desired vectors ($\tau = n$). By observing the structure of the vectors $\hat{\vect h}_{nm}(\tau)$ (see the graphical illustration for $M=3$ in Figure \ref{fig:alignment} on the top of the page), we notice that this alignment can be accomplished by setting
\begin{align}
      \frac{s_{nm'}(t,\tau)+h_{nm'}(t)}{h_{nm'}(\tau)}= \frac{s_{nm}(t,\tau)}{h_{nm}(\tau)}~,
\label{eqn:X_alignment_cond1}
\end{align}
for $\tau \neq n$, for all $t \in \{[N+1],\ldots,T\}$ and $m \in \{1,\ldots, M\}$, $m\neq m'$, where $m'$ is defined in (\ref{eqn:X_transmitters2}).
\subsection{Feasibility of Interference Alignment}
Equation (\ref{eqn:X_alignment_cond1}) can be rearranged by considering (\ref{eqn:entry_s}) to
\begin{subequations}
\begin{align}
\dsum_{j=1}^{J}{\mt{\vect g}_{nj}(t)\mat R_{j}(t,\tau) \Delta \vect f_{jnm}(\tau)} = b_{nm}(t,\tau) ~,
\label{eqn:X_alignment_cond2}
\end{align}
where 
\begin{align}
\Delta \vect f_{jnm}(\tau) = \left[h_{nm'}(\tau)\vect f_{jm}(\tau)-h_{nm}(\tau)\vect f_{jm'}(\tau)\right]
\end{align}
and 
\begin{align}
b_{nm}(t,\tau) = h_{nm'}(t)h_{nm}(\tau)~.
\end{align}
\end{subequations}
Exploiting \cite{charles}
\begin{align}
   \text{vec}(\mat A\mat C\mat B) = [\mt{\mat B} \otimes \mat A]\text{vec}(\mat C)~,
\end{align}
where $\mat A \in \setC^{l\times m}$, $\mat B \in \setC^{n \times p}$ and $\mat C \in \setC^{m \times n}$,
we can write equation (\ref{eqn:X_alignment_cond2}) as
\begin{subequations}
\begin{align}
    &\dsum_{j=1}^{J}{\mt{\vect v}_{jnm}(t,\tau)  \vect r_j(t,\tau)}= b_{nm}(t,\tau)~,
\label{eqn:X_alignment_cond3}
\end{align}
where 
\begin{align}
\mt{\vect v}_{jnm}(t,\tau) = \left[\Delta \mt{\vect f}_{jnm}(\tau)\otimes\mt{\vect g}_{nj}(t)\right]
\end{align}
and 
\begin{align}
\vect r_j(t,\tau) = \text{vec}(\mat R_{j}(t,\tau))~.
\end{align}
\end{subequations}
Now, let 
\begin{align}
\mt{\vect v}_{nm}(t,\tau) =\left[\mt{\vect v}_{1 nm}(t,\tau),\hdots, \mt{\vect v}_{J nm}(t,\tau) \right]
\end{align}
and 
\begin{align}
\vect r(t,\tau) =\text{vec}([\vect r_1(t,\tau),\hdots,\vect r_J(t,\tau)])~.
\end{align}
This allows us to rewrite (\ref{eqn:X_alignment_cond3}) as 
\begin{align}
      \mt{\vect v}_{nm}(t,\tau) \vect r(t,\tau) = b_{nm}(t,\tau)~.
\label{eqn:IC_lineqn}
\end{align}
By writing (\ref{eqn:IC_lineqn}) for all combinations of $n$ and $m$, with
\begin{subequations}\label{eqn:pairs_nm}
\begin{align}
    n &= 1,\hdots,N;~n\neq\tau,\\
    m &= 1,\hdots,M;~m\neq m'
\end{align}
\end{subequations}
we get $[(N-1)(M-1)]$ equations and we can write the system of linear equations for each pair of $(t \in \{[N+1],\ldots,T\},\tau \in \{1,\ldots,N\})$ using a matrix notation with $L$ unknowns
\begin{align}
\mat V(t,\tau) \vect r(t,\tau) = \vect b(t,\tau)~,
\label{eqn:X_lineqn2}
\end{align}
where 
\begin{align*}
&\mat V(t,\tau) \in \setC^{[N-1][M-1] \times L}~,\\
&\vect r(t,\tau) \in \setC^{L}~,\\
&\vect b(t,\tau) \in \setC^{[N-1][M-1]}~.
\end{align*}
Then, condition (\ref{alignment_condition}) is required to have a unique solution for the precoding matrices. Note that the matrix $\mat V(t,\tau)$ has full (row) rank almost surely, since all channel coefficients are drawn from a continuous distribution. For the equality sign of (\ref{alignment_condition}), we get a determined system of linear equations and the relays can obtain the precoding matrices by calculating $\mat V^{-1}(t,\tau)\vect b(t,\tau)$. The other case leads to an underdetermined system of linear equations and the relays use $\mat V^H(t,\tau)[\mat V(t,\tau)\mat V^H(t,\tau)]^{-1}\vect b(t,\tau)$, where $\mat V^H(t,\tau)$ denotes the conjugate transpose of the matrix $\mat V(t,\tau)$.

\subsection{Decoding}
User $n$ needs to decode the vector $\vect d_n$ which contains its desired symbols. We get the system of linear equations for user $n$ in matrix notation as follows:
\begin{align}
      \vect y_n = \hat{\mat H}_n \vect d~,
\label{eqn:X_decoding1}
\end{align}
with
\begin{align}
\hat{\mat H}_n = [\unit_T \otimes \mt{(\vect e_n^N)}] \tilde{\mat H} \mat U~,
\end{align}
\begin{align}
\mat U = \mleft{c} \unit_{NM} \\ \mt{(\ones_{N})} \otimes \vect e_1^M \mt{(\vect e_1^M)} \\ \vdots \\ \mt{(\ones_{N})} \otimes \vect e_{M-1}^M \mt{(\vect e_{M-1}^M)} \mright \in \setR^{TM \times NM}
\label{eqn:X_decoding2}
\end{align}
and $\vect d = \text{vec}([\vect d_1,\hdots, \vect d_N]) \in \setC^{NM}$, where $\ones_{K}$ is a $K$ dimensional column vector containing all ones. The matrix $\hat{\mat H}_n \in \setC^{T \times NM}$ is the $T$-symbol extended channel matrix received by Rx $n$. Due to the chosen precoding matrices, all interference vectors are aligned into an $[N-1]$ dimensional space and the desired signal vectors occupy the remaining $M$ dimensions whenever the condition
\begin{align*}
\dsum_{j=1}^{J}{L_{j}^{2}} \geq [N-1][M-1] 
\end{align*}
is fulfilled. This completes the proof of Theorem \ref{thm:theorem1}.

\begin{remark}
For the decoding of the symbol vector 
\begin{align}
\vect d_n =  \left[\mt{(\vect e_n^{N})} \otimes \unit_M \right] \vect d
\label{eqn:X_decoding5}
\end{align}
at Rx $n$ it is important that the desired vectors and the interference vectors are independent. This is ensured by the structure of the vectors given in (\ref{eqn:received signal_vector_usern3}). Since the matrix $\hat{\mat H}_n$ has full row rank almost surely, user $n$ is able to decode its symbols by considering (\ref{eqn:X_decoding1}), i.e.,
\begin{subequations}
\begin{align}
\vect d_n &= \left[\mt{(\vect e_n^{N})} \otimes \unit_M \right]\hat{\mat H}_n^+\vect y_n\label{eqn:X_decoding3}~,
\end{align}
with 
\begin{align}
\hat{\mat H}_{n}^{+} = \hat{\mat H}_{n}^{H}( \hat{\mat H}_n\hat{\mat H}_n^H)^{-1}~.
\label{eqn:X_decoding4}
\end{align}
\end{subequations}
The presented relay-aided transmit strategy allows us to communicate $NM$ symbols reliably in $T$ time slots, and therefore the DoF $\nicefrac{NM}{M+N-1}$ is achievable.
\end{remark}
\begin{remark}
The structured approach based on the Kronecker-product gives insights to the antenna configurations at the relays. We can see from equation (\ref{eqn:X_lineqn2}) that asymmetric relay antenna configurations are allowed as long as condition (\ref{alignment_condition}) is fulfilled. In that case, we get systems of linear equations with unique solutions.
\end{remark}
\begin{remark}
The transmit strategy of the transmitters in the time slots $t \in \{[N+1],\hdots,T \}$ is crucial to make the scheme work for time constant channels. We can see from equation (\ref{eqn:X_alignment_cond1}) that letting transmitter $m'$ be active, we have different alignment conditions for all $t \in \{[N+1],\hdots,T \}$. Even for time constant channels this leads to different precoding matrices for a $\tau \in \{1,\ldots,N\}$, since the right hand side of equation (\ref{eqn:X_alignment_cond2}) varies. This ensures that the $M$ desired signal vectors occupy an $M$ dimensional space. Note that the approach of \cite{30,31,32} is restricted to CSIT only, whereas our approach is applicable to a (static) X-network without CSIT. 
\end{remark}

\section{conclusion}\label{sec4}
In this paper, we have presented a systematic relay-aided communication strategy without CSIT for the $[M \times N]$-user X-network based on the Kronecker-product representation. We have shown that it is sufficient to install $J$ half-duplex relays with global CSI each equipped with $L_j$ antennas to achieve the maximum DoF of the network whenever condition (\ref{alignment_condition}) is fulfilled. Hence, the systematic approach results in generalized relay antenna configurations. Additionally, we have shown that the presented joint-beamforming strategy is applicable in static and slow fading environments. Finally, we emphasize that our approach unifies the analysis of time-varying and constant channels which might be useful for some practical scenarios.

\bibliographystyle{IEEEtran}
\bibliography{IEEEfull,references}

\end{document}